\documentclass[conference]{IEEEtran}
\IEEEoverridecommandlockouts

\usepackage{cite}
\usepackage{amsmath,amssymb,amsfonts}
\usepackage{algorithmic}
\usepackage{graphicx}
\usepackage{textcomp}
\usepackage{xcolor}
\def\BibTeX{{\rm B\kern-.05em{\sc i\kern-.025em b}\kern-.08em
    T\kern-.1667em\lower.7ex\hbox{E}\kern-.125emX}}

\usepackage{hyperref}

\hypersetup{
    pdfborder={0 0 0}
}

\PassOptionsToPackage{usenames,dvipsnames}{xcolor}

\usepackage[switch,columnwise]{lineno}
\usepackage{tikz}
\usepackage{pifont}

\usepackage{filecontents}

\usepackage{graphicx}
\usepackage{float}
\usepackage{booktabs}
\usepackage[usenames,dvipsnames]{xcolor}

\usepackage{xurl,hyperref}

\newcommand{\TROJAN}{\mbox{\textsc{Surf}}}

\usepackage{tabularx, multirow, graphicx}

\usepackage{listings}

\usepackage{dirtytalk}

\usepackage[symbol]{footmisc}

\newcommand*\circled[1]{\tikz[baseline=(char.base)]{
            \node[shape=circle,draw,inner sep=1.2pt] (char) {#1};}}

\usepackage[numbers,sort&compress]{natbib}

\begin{document}
\bstctlcite{IEEEexample:BSTcontrol}

\title{Exploiting Software-level Abstractions To Support Practical Hardware Trojan Attacks
}

\author{
\IEEEauthorblockN{
Athanasios Moschos \qquad
Kevin Valakuzhy \qquad
Georgios Kokolakis
}

\IEEEauthorblockN{
Fabian Monrose \qquad
Angelos D. Keromytis
}

\IEEEauthorblockA{
\textit{Georgia Institute of Technology}\\
Atlanta, Georgia, USA\\
\{amoschos, kevinv, gkokolakis6, angelos\}@gatech.edu,
fabian@ece.gatech.edu
}
}

\maketitle
\thispagestyle{plain}
\pagestyle{plain}

\begin{abstract}
    Hardware trojan (HT) attacks against CPUs typically assume threat scenarios where an attacker targeting a system with a trojanized CPU is able to execute arbitrary code (\textit{i.e.} machine-level instructions) to reliably interact with the implanted trojan.
On end-user devices (\textit{i.e.,} mobiles, laptops), achieving arbitrary code execution in practice requires software exploits tailored to each specific target.
Such strong adversarial premises reduce the generality of existing threat models casting doubt on CPU trojan attacks as a pragmatic threat vector.

\par To push the envelope on HT attacks against client devices, we introduce the \TROJAN~class of CPU-trojans that can be activated without arbitrary code execution.
Our key insight is that integer operations expressed in a high-level language can be mapped to microarchitectural side-effects distinguishable by a \TROJAN~trigger circuit.
This observation unlocks HT activation via runtime engines, constrained environments executing untrusted high-level code. 
We demonstrate a \TROJAN~trojan inside a RISC-V processor and exploit JavaScript-level memory indexing operations inside Google's V8 engine to perform a code injection attack.
Importantly, we show that \TROJAN~trojans remain effective across multiple JavaScript engine versions, enabling long-term compromise of endpoint devices.
To facilitate research, we open-source \TROJAN's design and supporting software.
\end{abstract}


\section{Introduction}
\label{sec:introduction}
    Hardware trojans are often associated with offensive campaigns against high-value military systems~\cite{IEEEX:ADE08} and are typically attributed to state-level actors.
The offensive security community, however, recognizes that similarly valuable targets include personnel within accredited organizations (\textit{e.g.,} diplomats) and security-critical corporations such as cybersecurity firms.
The disclosure of recent activities like Operation Triangulation~\cite{web_LB2023, web_GD2023} demonstrates how third-party intellectual property (3PIP) in commercial \textit{endpoint devices} can be weaponized against special interest individuals who own them.
Albeit not a hardware trojan per se, Operation Triangulation is the closest instance we have of a large-scale offensive campaign designed to exploit a silicon feature in a third-party processor to gain a persistent adversarial foothold on end-user devices incorporating it.
More broadly, the incident points to an important realization: advances in the defensive posture of modern endpoint devices have raised the bar beyond what pure software exploits can reliably achieve, increasingly pushing sophisticated attacks toward the exploitation of hardware logic.
This real-world incident lends credence to the research community's early concerns about the potential of processor hardware hosting powerful trojan implementations~\cite{DBLP:TM14, DBLP:YHDAS16, DBLP:DeKNG20, ACM:VGHGSPJR22, KDJC23, MAMFKAD24, CCSHAACLT24}.

\par The interplay between hardware and software in modern processors is, nonetheless, complex. 
Binary-level code execution becomes essential to leverage hardware features buried deep in the microarchitecture.
As a case in point, the exploitation chain of Operation Triangulation starts with a zero-day exploitation~\cite{web_LB2023} which leads to \textit{arbitrary code execution} on the victim devices.
Likewise, typical hardware trojan threat models~\cite{DBLP:YHDAS16, DBLP:DeKNG20, KDJC23, MAMFKAD24} assume attackers can \textit{activate trojans} via arbitrary machine-level instructions.
This ability, however, is not the attack outcome but rather a prerequisite necessary to jump-start the offensive campaign.
While easy in theory, practical realization of arbitrary code execution requires chaining multiple software-level flaws into viable gadget chains~\cite{SKZMFDLDALCSAR2013}.
Yet, operating system (OS) diversity, coupled with heterogeneous software stacks and security configurations, renders this realization a daunting task.
Thus, the common assumption that attackers have readily available fine-grained execution rights on target devices, despite the difficulty of obtaining them at scale, can create a \textit{false sense of security about the relevance of hardware trojan threats to end-user devices}.

\par We posit that attackers can embed trojans in client-device CPUs that obviate the need for arbitrary code execution during activation, minimizing reliance on additional device or software stack vulnerabilities.
Our work explores this alternative by leveraging \textit{execution engines of runtime systems} for trojan triggering.
Runtime engines are integral to applications processing untrusted content (\textit{e.g.,} browsers, document readers), exhibit stable behavior across diverse hardware platforms and can therefore serve as reliable activation vectors.
Building on these observations, we introduce \textit{\TROJAN-trigger circuits}, which exploit predictable hardware manifestations of software-level computations to activate malicious payloads.
Specifically, we show that an attacker can leverage a runtime engine to stimulate an embedded \TROJAN~trojan, which then locates and executes malicious machine-level code injected in the memory of an otherwise bug-free runtime system. 
In summary:
\begin{itemize}
    \item We introduce \TROJAN-trigger circuits and explore the practicality of stimulating them via JavaScript-level integer operations executed inside the popular V8 engine.
    \item We prototype a \TROJAN~trojan inside a Linux-capable RISC-V processor on FPGA and exploit the effective address calculation of JS-level memory indexing operations to perform a code injection attack.
    \item Using different V8 engine releases, we evaluate the robustness of \TROJAN-triggers in terms of \emph{(i)} resisting possible detection efforts and \emph{(ii)} operating under diverse workload conditions, among an evolving software stack.
\end{itemize}
Altogether, our practical trojan activation approach highlights the threat posed by HTs to modern endpoint devices.

\section{Background}
\label{sec:background}
    \subsection{CPU Hardware Trojan Attacks}
\label{sec:hw_trojans}
    \par
\begin{figure}[t]
    \centering
    \includegraphics[width=0.8\columnwidth]{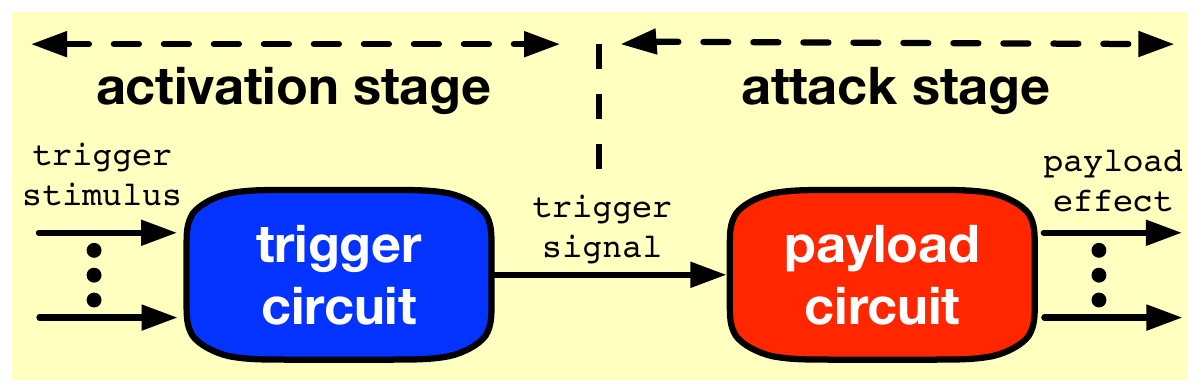}
    \caption{Operational stages of a hardware trojan attack.}
    \label{fig:trojan_attack_stages}
\end{figure}

 Hardware trojans comprise a trigger and a payload circuit, with HT offensives reflecting two distinct operational stages, activation and attack, illustrated in Figure~\ref{fig:trojan_attack_stages}. 
During the activation stage, the trigger monitors for a special condition to enable the payload, while in the attack stage the payload manifests the malicious behavior.

\par A trojan implementation is \textit{practical} only if it can be reliably activated at the attacker's behest.
In CPU-based hardware trojans, activation---that is, the generation of the trigger signal---typically depends on the execution of attacker-controlled software on the target system~\cite{DBLP:YHDAS16, ZJZYLHJJ20, DBLP:DeKNG20, MAMFKAD24} or on the injection of adversarial data that is observable by the malicious hardware~\cite{DBLP:KTCGJZ08, MAVKKAD2022}.
Crucially, establishing that a trojan is controllable requires attackers to reason explicitly about their assumed level of access to a victim system equipped with an infected CPU.
For the remainder of this paper, we refer to this reasoning as the \textit{trojan activation model}.

\par The activation model, typically part of the attack's threat model, delineates not just the trigger stimulus, but also the capabilities an adversary must have to successfully start a hardware trojan offensive.
Specifying this model enables stakeholders to assess if they fall within a trojan's effective threat surface and develop defenses that disrupt its activation. 

\subsection{Trojan Activation Models}
\label{sec:activation_models}
    \par The efforts of~\citet{CCSHAACLT24} and~\citet{MAMFKAD24} survey the main research on processor HTs.
Since the methodologies and scopes of these studies vary widely, we filter them to create a body of works that explicitly discuss details about the underlying activation model of each trojan.
First, we exclude works that do not analyze HT controllability or activation conditions~\cite{ZJXQ13, ACM:AHTPSPGS, ACM:SPMGFSTRD23, RJVVKR15, LDZQZDLLHJYYZY22, FNKSCKTT15, HMCJCPBSHT2022}.
Cryptographic accelerators~\cite{JYKNMY09, MDGHWSM20, HWZLAABJHBTYKR17} are monolithic circuits with no OS support and therefore outside our scope.
On these principles, we exclude the work of~\citet{DBLP:HS21} on the Riscy microcontroller.
The resulting papers considered in our analysis are summarized in Table~\ref{table:literature}, along with the adversarial \textit{ability} required to generate the trigger circuit \textit{stimulus} necessary for the trojan activation.

\par Trigger circuits are often constructed using wires with low switching activity~\cite{DBLP:YHDAS16, DBLP:KAFP19, ACM:VGHGSPJR22}.
Stimulating such wires typically requires carefully crafted instruction sequences that may \emph{(i)} enforce precise ordering, \emph{(ii)} access special-purpose registers, \emph{(iii)} use specific immediates or offsets, \emph{(iv)} select particular registers or \emph{(v)} combine legal with illegal instruction encodings.
The latter approach is also used as a stimulus in~\citet{ZJZYLHJJ20}.
Additional trigger design considerations may necessitate high-frequency repetition of these instruction streams within short time windows~\cite{DBLP:TM14, DBLP:YHDAS16, DBLP:KAFP19}.
To meet these constraints, activation models in prior work assume attackers capable of executing arbitrary binary-level code on the victim systems.
Similar assumptions underlie alternative activation strategies, including trojans triggered via a wall-clock timer configured through a debug logic backdoor (\textit{e.g.,} in the scan chains)~\cite{KMHHCMLKJ19} and those relying on the virtual address (VA) reconnaissance of the victim software stack~\cite{KDJC23}.
Likewise, approaches based on register manipulation~\cite{MAMFKAD24} or arbitrary memory operations~\cite{DBLP:DeKNG20, DBLP:KTCGJZ08} assume attackers can first deploy their own process on the target system at the activation stage.
In security parlance, this attacker capability fueling the above activation models is commonly referred to as \textit{arbitrary code execution}.   
Only a minority of trojans feature activation conditions that do not require this capability~\cite{KDJC23, DBLP:KTCGJZ08, MAVKKAD2022}.

\begin{table}[t]
\footnotesize
\centering
\caption{Trojan Activation Models in CPU Hardware Trojan Attacks}
\scalebox{0.8}{
\label{table:literature}
\begin{tabular}{*{3}{c}}
    \toprule
    \textbf{Publication}& \textbf{Activation Stage Ability}& \textbf{Trigger Stimulus}\\
    \midrule
    \cite{DBLP:TM14}& \shortstack{a) - \\ b) arbitrary code execution}& \shortstack{a) no trigger; always-on design \\ b) repetitive instruction pattern}\\
    \hline
    \cite{ACM:VGHGSPJR22}& arbitrary code execution& precise instruction pattern\\
    \hline
    \cite{DBLP:KAFP19}& arbitrary code execution& repetitive instruction pattern\\
    \hline
    \cite{DBLP:YHDAS16}& arbitrary code execution& repetitive instruction pattern\\
    \hline
    \cite{ZJZYLHJJ20}& arbitrary code execution& mix of legal \& illegal instructions\\
    \hline
    \cite{KDJC23}& \shortstack{a) arbitrary code execution\\ b) control a type-safe interface}& \shortstack{a) virtual address of a bounds check\\ b) instructions of a bounds check}\\
    \hline
    \cite{MAMFKAD24}& \shortstack{a) arbitrary code execution\\ b) arbitrary code execution}& \shortstack{manipulation of targeted registers\\ integer operations\footnotemark[1]}\\
    \hline
    \cite{DBLP:DeKNG20}& arbitrary code execution& memory operations\\
    \hline
    \cite{DBLP:KTCGJZ08}& \shortstack{a) arbitrary code execution\\ b) influence over D\$}& \shortstack{a) memory operations\\ b) network packets}\\
    \hline
    \cite{KMHHCMLKJ19}& arbitrary code execution& POSIX clock time event\\
    \hline
    \cite{MAVKKAD2022}& compiler allocation of targeted registers& network packets\\
    \midrule
    \TROJAN& high-level code execution& integer operations\\
    \bottomrule
\end{tabular}
}
\end{table}
\footnotetext[1]{Authors only consider use of assembly to form the integer operations.}

\par Frequently, the focal point in offensive research of hardware trojan attacks is either the hardware-level design novelty, the insertion method, or the attack stage functionality, with each serving to highlight the potential impact of this clandestine threat.
As a result, authors often gloss over the preconditions that, in practice, define a favorable setting for the activation stage of an offensive.
In prior literature, such settings are typically reduced to overly simplified assumptions in which adversaries either possess a user account on the victim system~\cite{DBLP:YHDAS16, ZJZYLHJJ20, DBLP:DeKNG20}, operate within a virtualized environment~\cite{DBLP:YHDAS16, KDJC23}, or can execute malicious software~\cite{DBLP:KTCGJZ08, KMHHCMLKJ19, DBLP:DeKNG20}.

\par In reality, though, obtaining user-level access on large populations of endpoint devices is impractical.
Establishing a foothold on end-user devices becomes a case-by-case task (\textit{e.g.,} with victim-specific software exploits), as their diverse operating systems and software stacks lead to heterogeneous security configurations.
Additionally, frequent software updates or patching can shorten or eliminate such windows of opportunity to leverage a trojan.
In light of these considerations, we argue that activation models dependent on arbitrary code execution can \textit{significantly diminish the credibility of hardware trojans as a viable threat against endpoint devices}.

\subsection{Trojan Activation Via Runtime Engines}
\label{sec:remote_aa}
    \par Given these constraints, trojan activation on endpoint devices calls for strategies independent of arbitrary code execution.
Instead, achieving remote interaction with HTs at scale demands ubiquitous software platforms on the device side that expose interfaces capable of influencing the underlying hardware behavior from a distance (\textit{e.g.,} over the network).
To offer consistent user experience across different hardware platforms, commercial applications rely heavily on the use of \textit{runtime systems}.
Hence, we consider runtime environments in modern software stacks as a practical substrate that combines both ubiquity and remotely accessible interfaces.

\par Runtime systems incorporate execution engines to enable seamless software deployment across devices and operating systems.
Acting as intermediates between software and hardware, they offer the following benefits:
\begin{itemize}
    \item Code portability, as execution engines translate high-level code into (local system) machine-level instructions, enabling cross-platform application compatibility.
    \item Execution consistency, ensuring predictable application behavior across heterogeneous systems.
\end{itemize}
In essence, runtime engines \emph{(i)} are native components of the system's software stack (\textit{e.g.,} not installed by attackers), \emph{(ii)} are common in applications that that process attacker-controlled content (\textit{e.g.,} web browsers and document readers), and \emph{(iii)} expose code structures (\textit{e.g.,} functions) with stable behavior across systems.
Consequently, we argue that runtime execution engines constitute promising access vectors for triggering hardware trojans, as they can \textit{decouple activation from the preconditions of arbitrary code execution}.

\paragraph*{\textbf{Contrast to arbitrary code execution}}
Adversarial access to runtime engines is substantially easier to obtain as modern endpoint devices intentionally expose environments designed to execute untrusted code. 
Reaching these execution contexts typically requires only content delivery (\textit{e.g.,} visiting a webpage, loading remote user interface content, or opening a document).
Importantly, runtime engines by design preclude processing of machine-level instructions.
Instead, the code is expressed in high-level abstractions (\textit{e.g.,} objects, functions) that lack direct machine code equivalents, a disconnect commonly referred to as the \say{semantic gap}.

\par To bridge this gap, runtime engines use interpreters or just-in-time compilers to translate the high-level code into machine instructions.
These intermediate software layers---and not the adversaries---ultimately determine the instruction streams executed within the CPU.
Consequently, runtime engines deprive user control over \emph{(i)} the machine instructions emitted, \emph{(ii)} their ordering, \emph{(iii)} the registers or immediates they operate on, or \emph{(iv)} their precise timing and frequency.
These are the bedrock of several prior work activation models~\cite{DBLP:TM14, ACM:VGHGSPJR22, DBLP:KAFP19, DBLP:YHDAS16, MAMFKAD24}.
Such limitations are not favorable either to activation models founded on malicious software execution on the target system~\cite{DBLP:KTCGJZ08, KMHHCMLKJ19, DBLP:DeKNG20} or hunting for virtual addresses via cache side-channels~\cite{KDJC23}.
Lastly, the runtime compilers and interpreters cannot issue illegal instruction encodings featured in the stimuli of~\cite{ZJZYLHJJ20, DBLP:KAFP19}.

\par Taken together, the above factors make runtime engines not applicable for the activation models seen in prior work.
In the following sections we present a novel approach, applicable to runtime engines, that leverages \textit{integer operations} expressed in high-level code for activating hardware trojans.

\section{Threat Model}
\label{sec:threat_model}
    \par Our work adopts the threat model of a 3PIP vendor attack~\cite{Xue2020TenYO, AYKF2008, TMKF2010}, where a trojan is embedded within a third-party processor IP integrated by the design house or SoC developer.
Vendors typically sell 3PIPs in a black-box form, either as a netlist description or a tape-out-ready layout representation, to protect their proprietary designs, maintain their competitive advantage and preserve profitability.
Due to this black-box nature~\cite{AD2006}, a malicious IP will be incorporated as is with the rest of the chip, therefore increasing the likelihood of a stealthy attack.
We deem this a plausible attack route, given the widespread integration of 3PIPs in modern SoCs~\cite{Xue2020TenYO, web_PE2024}.

\par We assume the chip with the malicious processor has passed any security-relevant industry standard checks and launched on the market as a commercial off-the-shelf (COTS) product.
This is considered feasible as existing verification methods and applicable detection techniques~\cite{Xue2020TenYO} cannot guarantee the non-existence of additional (malicious) behaviors, especially if the nominal functional design specification is unmodified~\cite{TNGKCMM2014}.
To the best of our knowledge, trojan detection methods are not yet an industry standard and thus, there is no guarantee that IPs are tested against them before integration.

\par Operationally, we consider an attacker who can remotely interface with a trojan-infected endpoint device.
The victim system offers no exploitable hardware or software vulnerabilities an attacker can leverage during \TROJAN's activation stage, and runs an application with a bug-free execution environment.
A canonical example is a web browser, which provides both a network-facing interface and a runtime execution environment (\textit{e.g.,} a JavaScript engine); however, other applications exposing runtime environments, such as document viewers, also fit our model.
By executing untrusted code (\textit{e.g.,} via phishing links, malicious advertisements, compromised legitimate sites or malicious documents), the attacker influences the machine-level instructions executed by the runtime engine.
Specifically, the adversary can induce computations expressed in high-level code (\textit{e.g.,} JavaScript).
Our trigger mechanism leverages the predictable manifestation of these JavaScript-level computations in the underlying CPU hardware to activate the trojan payload.
From an attacker's perspective, the JavaScript path is preferable over WebAssembly execution, since the latter is not available in document readers.
Moreover, in the context of browsers, WebAssembly is widely regarded as a potential red flag~\cite{MMWCJMRK2019, HHMD2024} that can expose an offensive campaign.



\section{\TROJAN~Trigger Circuits}
\label{sec:surf_trojans}
    The lack of precise control over the machine code emitted by runtime systems complicates their use for trojan activation. 
Consequently, attackers must rely on alternative sources of determinism to interact with HTs.
To this end, we observe that integer operations (e.g., arithmetic and logical) expressed in high-level code exhibit \textit{predictable and observable microarchitectural effects}, suitable for stimulating \TROJAN-trigger circuits.

\subsection{Trigger Stimulus}
\label{sec:surf_operations}
    \paragraph*{\textbf{Integer operations}}
In general-purpose microarchitectures, integer operations such as addition, subtraction, and bitwise logic (\textit{e.g.,} AND, OR, XOR) constitute fundamental architectural primitives---meaning built-in, lowest-level operations that the hardware directly supports.
In high-level languages, these operations are typically expressed with two explicit operands. 
Runtime interpreters and compilers preserve their semantics when lowering them to architectural instructions (\textit{e.g.,} \texttt{add}, \texttt{sub}, \texttt{and}, \texttt{or}, \texttt{xor} for RV64I ISA) which operate on the values held in their source registers.
Ultimately, by selecting the operands of high-level integer operations, attackers can influence the source register values of the machine instructions these operations map to.
The \TROJAN-trigger circuits can identify the manifestation of these instructions in the processor's hardware by tracing their operand values.

\paragraph*{\textbf{Integer Operation Pattern}}
A \TROJAN~trojan is activated when a key, a sequence of integer operations (IOs) with attacker-controlled operands, is executed in the processor.
We refer to this activation key as an \textit{Integer Operation Pattern} or IOP.
An IOP is characterized by properties that govern how the trigger stimulus is constructed and identified.
First, it comprises an arbitrarily long sequence of integer operations.
Second, each instruction's operand pair is \textit{asymmetric}: one operand, referred to as \textit{free}, is unconstrained and may take any value, while the other, named \textit{static}, is restricted to a predefined set of values decided at \textit{design time}.
Importantly, although the free operand may assume any value \textit{at runtime}, it must remain constant across the entire IOP sequence to form a valid key.
This structure increases triggering flexibility without compromising trigger stealthiness, as we show in Section~\ref{sec:surf_detection_resilience}.
Third, the sequence itself is fixed: both the static operand set and the order in which these values need to appear are decided at design time.
While the corresponding operations must occur in this order, \textit{they need not be contiguous in the instruction stream}.
Therefore, unrelated computations may interleave with IOP elements without preventing successful key recognition.
Finally, the attacker may choose at runtime any mix of integer operations to form the IOP sequence.

\subsection{Trigger Circuit Design}
\label{sec:surf_trigger_design}
    \par
\begin{figure}[t]
    \centering
    \includegraphics[width=0.9\columnwidth]{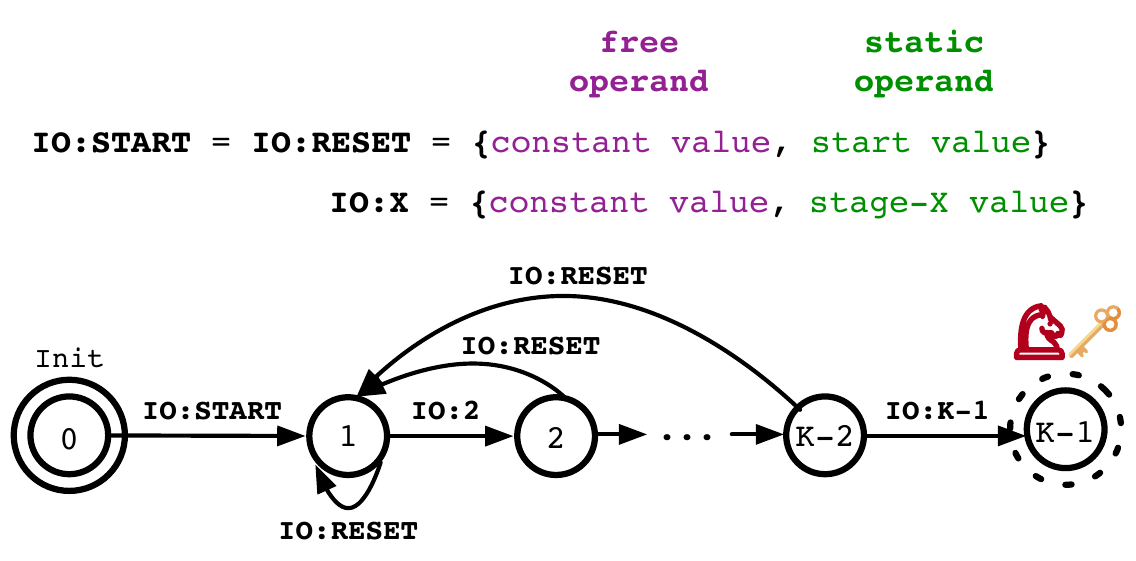}
    \caption{Finite state machine of a SURF trigger circuit.}
    \label{fig:IOP_Trigger_Design}
\end{figure}

 An IOP is identified by the trigger circuit depicted in Figure~\ref{fig:IOP_Trigger_Design}, implemented as a finite state machine (FSM).
The designated integer operation to initiate an IOP sequence is denoted as \texttt{IO:START}. 
In this operation, the \texttt{start value} inside the static operand signals the FSM to latch onto the \texttt{constant value} carried by the free operand.
To advance through subsequent states, the trigger mechanism monitors the instruction stream for integer operations whose source registers simultaneously satisfy \emph{(i)} the latched constant value and \emph{(ii)} the expected static value for the next state.
If these conditions are not met, the FSM remains at its current state; benign integer operations may execute without affecting key recognition, in accordance with the IOP properties.
Upon observing the final operation contributing to the IOP, the trigger signal is asserted and the activation stage shown in Figure~\ref{fig:trojan_attack_stages} is concluded.

\par During the CPU operation, benign instruction sequences that partially satisfy the IOP criteria can transiently advance the FSM.
To prevent such scenarios from blocking a subsequent attacker-directed activation attempt, we integrate a reset mechanism that re-initializes the FSM upon the arrival of an operation satisfying the \texttt{IO:START} characteristics.
This operation, denoted as \texttt{IO:RESET} in Figure~\ref{fig:IOP_Trigger_Design}, re-latches the fresh value of the free operand, overriding the previously captured value.
So \texttt{IO:RESET} and \texttt{IO:START} are functionally equivalent.

\section{\TROJAN-Enabled Code Injection Attacks}
\label{sec:end_to_end_attack}
    \par Our goal is to demonstrate that \TROJAN~trojans not only enable a more relaxed activation model but can also establish powerful footholds within their host devices.
To this end, we craft a \TROJAN-based trojan that locates and executes machine code injected into the memory of a running process.
We evaluate our trojan implementation on the V8 platform, Google's open-source, high-performance JavaScript engine, that empowers browsers like Chrome, Brave, Opera and Microsoft Edge.
Given Chrome's dominant market share~\cite{ArticleVJ2024}, our choice ensures broad practical relevance.
This attack utilizes untrusted JavaScript executed within the V8 environment to enable both the activation stage and the attack stage depicted in Figure~\ref{fig:trojan_attack_stages}.

\paragraph*{\textbf{Attack Impact}}
A successful code injection attack yields the ability to execute arbitrary code inside Chrome's renderer process.
This constitutes a critical security exposure~\cite{cve-2026-11645} that, when paired with independent vulnerabilities or policy flaws (\textit{e.g.,} a browser sandbox escape) in other runtime components or the operating system kernel, can lead to total compromise of the endpoint device.
Crucially, this capability enables the deployment of most CPU trojans in Table~\ref{table:literature}, whose operation assumes arbitrary code execution on the target device.

\begin{figure}[t]
    \centering
    \includegraphics[width=0.75\columnwidth]{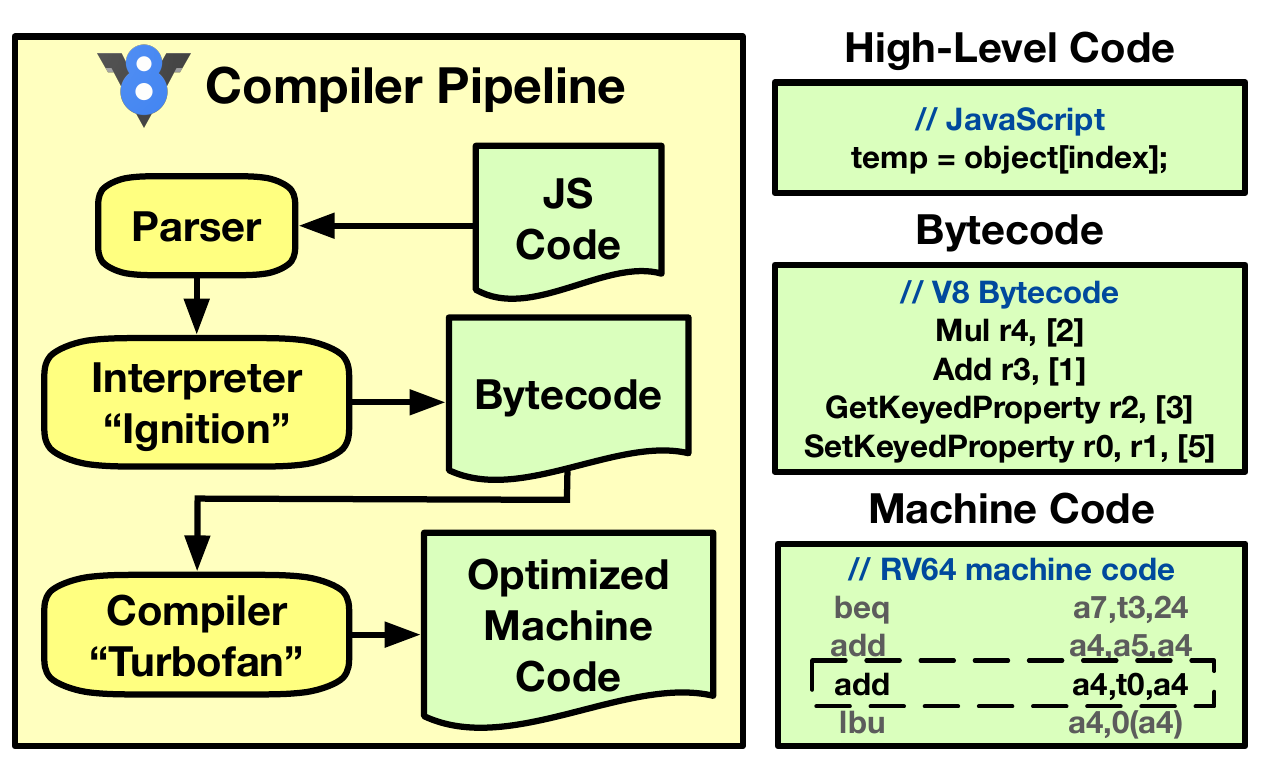}
    \caption{Overview of the V8 engine compiler pipeline and the representation of JavaScript code within it. Highlighted is the machine instruction responsible for the effective address calculation contributing to the IOP generation.}
    \label{fig:v8_compiler_pipeline_code_comb}
\end{figure}

\paragraph*{\textbf{The V8 engine}} 
Execution engines use interpreters and compilers to dynamically translate high-level code into machine instructions at runtime and optimize its execution.
V8 engine's execution pipeline primarily relies on two language processors, the Ignition interpreter and the Turbofan compiler~\cite{V8IgniTurbo}.
As shown in Figure~\ref{fig:v8_compiler_pipeline_code_comb}, JavaScript code is initially parsed and forwarded to Ignition which generates and intermediate bytecode representation and dispatches for execution the corresponding pre-compiled machine-code handler of each bytecode instruction.
In parallel, Ignition profiles execution to identify frequently executed bytecode regions, which are passed to the just-in-time TurboFan compiler, for translation into optimized machine code.
Figure~\ref{fig:v8_compiler_pipeline_code_comb} illustrates the resulting code representation at each stage of this pipeline.

\subsection{Code Injection Attack Stages}
\label{sec:code_injection_attack}
    \paragraph*{\textbf{Activation Stage}}
During this stage, the attacker-controlled JavaScript executed by the runtime engine serves two purposes.
First, it injects a software payload into the V8 engine’s virtual memory as a JavaScript object (\textit{e.g.,} an array) containing native machine instructions in byte form.
Second, it performs integer computations that serve as trigger stimuli, forming the IOP trojan activation key.

\paragraph*{\textbf{Attack Stage}}
Upon activation, the trojan hijacks the control flow of the V8 process and \textit{redirects execution} from the JS code to the byte contents of the injected object.
The injected native instructions execute until control is restored to the original JS code, which concludes the attack by issuing an IOP to deactivate the trojan.
Fundamentally, successful redirection first requires recovering the virtual address of the injected software payload.
Next, we describe how the \TROJAN-trigger mechanism enables this recovery, followed by the implementation of the trigger and payload circuits.


\begin{figure*}[t]
    \centering
    \includegraphics[width=0.8\textwidth]{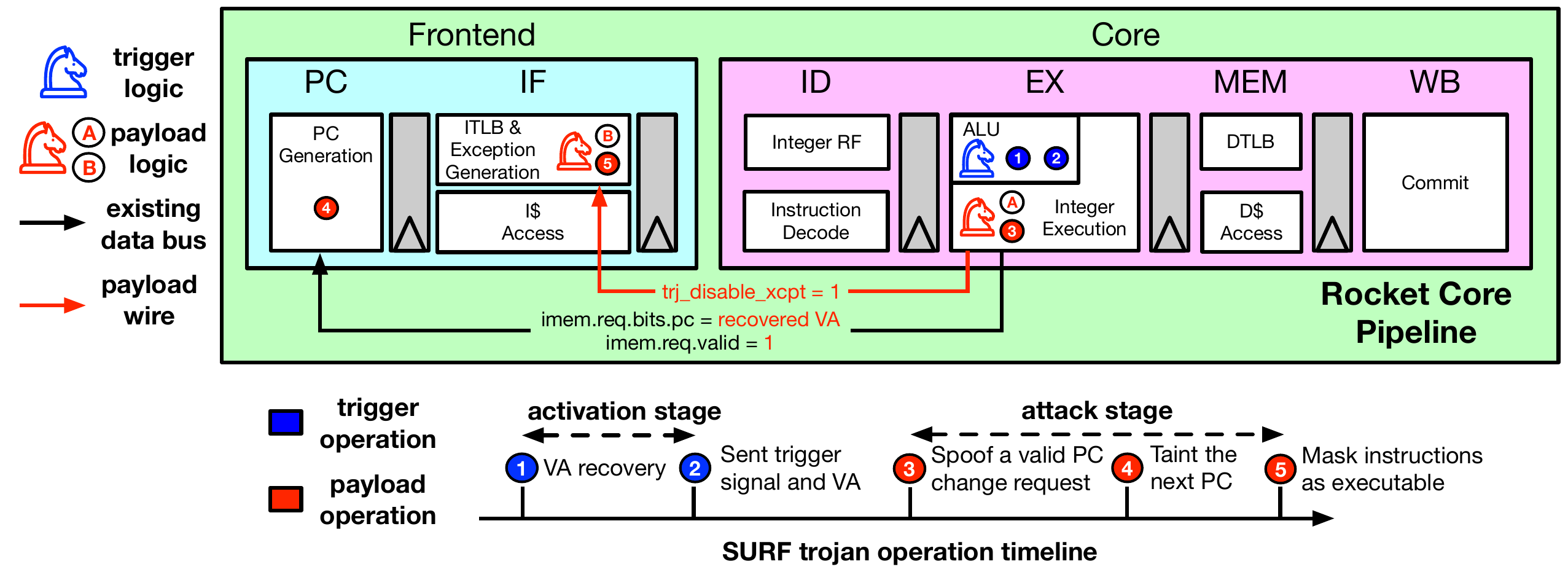}
    \caption{The Rocket Core pipeline with a \TROJAN~trojan for code injection attacks. The trojan horses signify logic modifications, while the enumerated circles signify operational activity related either with the trigger or the payload logic.}
    \label{fig:rocket_pipeline}
\end{figure*}

\subsection{V8 Engine Memory Reconnaissance}
\label{sec:memory_reco}
    \par To enable code portability and enhanced security, JavaScript abstracts direct memory management and does not expose virtual or physical addresses to the programmer.
Memory allocation is handled by the V8 engine backend, while operating-system-enforced address space layout randomization (ASLR) randomizes its virtual memory layout.
As a result, the adversary is oblivious to the virtual address of the JS object carrying the software payload.

\par Nevertheless, at the JavaScript-level, the adversary can access the object through \textit{memory indexing} operations (\textit{e.g.,} reading or writing its elements).
At the machine-code level, such operations require computing the \textit{effective address} of the accessed element by adding the object’s randomized base address to the attacker-chosen index.
Consequently, a sequence of these calculations can satisfy the IOP properties outlined in Section~\ref{sec:surf_operations}: the randomized base address serves as the free operand, while the selected indices correspond to the values of the static operand drawn from a predefined set.
More importantly, repeated accesses to the same object preserve the base address in the free operand, regardless of the index used.

\par This high-to-low code transformation contributing to the \TROJAN~key generation is illustrated in Figure~\ref{fig:v8_compiler_pipeline_code_comb}.
A byte read from a JavaScript object is lowered to RV64 instructions, where the source registers of an \texttt{add} instruction expose the object's raw base pointer and the index used to compute the effective-VA.
This raw base-pointer corresponds to the virtual address in the V8 address space, prior to its translation into a physical address by the memory management unit.

\par Hence, the trigger circuit presented in Section~\ref{sec:surf_trigger_design} can recover the software payload’s virtual address, \textit{effectively bypassing ASLR}, and facilitate control-flow hijacking.
Importantly, \TROJAN~ achieves this entirely within the processor, without external memory disclosures or software-level information leaks.
The trigger circuit recovers the address as a byproduct of the activation procedure, eliminating the need for auxiliary disclosure channels.
In contrast, the ROP-based attack in~\cite{KDJC23} requires arbitrary code execution on the victim system to break ASLR via cache-based side channels, before trojan activation.

\subsection{\TROJAN~Trojan Implementation}
\label{sec:surf_implementation}
    \par We evaluate our \TROJAN~trojan inside the 5-stage, in-order Rocket Core microarchitecture~\cite{rocketchip2016}, which reflects the fundamental execution and memory structure of a Linux-capable general-purpose core implementing a modern instruction set.

\paragraph*{\textbf{Integer Operand Visibility}}
As discussed in Section~\ref{sec:surf_trigger_design}, the trigger circuit requires visibility into both operands of an integer computation.
The Rocket Core pipeline abstraction in Figure~\ref{fig:rocket_pipeline}---floating-point portion is omitted for simplicity---illustrates such candidate monitoring locations.
Operands originating from the runtime engine’s virtual memory are loaded into architectural registers and subsequently read from the register file by integer instructions that reach the decode stage (ID).
Before reaching the execution stage (EX) and the inputs of the arithmetic logic unit (ALU), they pass through the bypass network to resolve data hazards and get latched into pipeline registers.
Thus, multiple candidate locations exist between the ID and EX stages expose operand values to detect the IOP activation pattern.

\paragraph*{\textbf{Trigger Circuit}}
We implement our \TROJAN-trigger inside the arithmetic logic unit (\textcolor{blue}{blue trojan horse} symbol in Figure~\ref{fig:rocket_pipeline}). 
Compared to earlier pipeline locations, the ALU has visibility to source registers of integer instructions only, reducing the volume of data variance in the trigger stimuli.
The circuit implements the state machine depicted in Figure~\ref{fig:IOP_Trigger_Design} and passively monitors the ALU's two 64-bit inputs.

\par The trigger logic implementation addresses two challenges: \emph{(i)} reliably distinguishing IOP stimuli from benign integer computations, and \emph{(ii)} the lack of control, at high-level code, over which ALU input receives the injected object's address versus its index.
For the first, we exploit the expected form of operand values (\textit{i.e.,} a virtual address) involved in our attack scenario, to isolate the trigger stimuli.
Rocket Core uses a page-based 39-bit virtual address space; therefore the FSM's free operand in Figure~\ref{fig:IOP_Trigger_Design} is restricted to canonical user-space addresses (\textit{e.g.,} values with bits \texttt{[63:39]} equal to zero).
We further require at least one bit within a predefined subset of the lower 32 bits to be set, filtering out very small integer values.
For the static operand, values must match the indices selected during the trojan design phase, with their upper bits set to zero.
Together, these constraints distinguish the trigger stimuli from varying traffic generated by benign computations.

\par Turning to the second challenge, high-level code cannot control which ALU input receives the address and which the index value.
However, we observe that runtime interpreters and compilers rely on fixed, pre-compiled machine-code stubs (\textit{e.g.,} V8 Built-in functions), yielding consistent selection of ALU inputs across repeated executions of an integer operation.
We therefore duplicate the FSM with swapped ALU inputs, reliably identifying an IOP irrespective of its operand ordering.

\paragraph*{\textbf{Payload Circuit}}
A successful attack requires redirecting the PC outside the normal V8 execution flow, and permitting execution at the target address.
Our hardware payload achieves both by \say{tainting} the next-PC generation with the JavaScript object's address while suppressing exceptions that would prevent its execution.
We realize this by modifying the core's EX stage logic and the frontend's exception logic, shown by the \textcolor{red}{red trojan horse} symbols in Figure~\ref{fig:rocket_pipeline}.

\par Rocket Core's execution stage issues PC-change requests to the frontend through a dedicated bus in response to control-flow changes, exceptions or stalls.
Our payload modifies this logic---marked as \circled{A}---to spoof a control flow change, by driving the target address on the (\texttt{imem.req.bits.pc}) bus and asserting the \texttt{imem.req.valid} to indicate a valid PC-change request.
The payload logic then protects this request against microarchitectural events that could invalidate the redirection.
For example, a subsequent memory instruction incurring a cache miss will be replayed, reverting execution from the injected instruction stream.
To prevent this, payload logic suppresses subsequent PC-change requests by holding the \texttt{imem.req.valid} low until the first instruction fetched from the redirected stream executes.

\par In the frontend, the multiplexer marked \circled{B} is controlled by the \texttt{trj\_disable\_xcpt} trojan wire, which suppresses instruction page faults and access exceptions upon trojan activation. 
This enables the execution of injected instruction streams in protected or non-executable memory.





\subsection{\TROJAN~Trojan Operation}
\label{sec:surf_operation}
    \par We prototype our Rocket Core \TROJAN~trojan on a Genesys 2 FPGA board to evaluate the practicality of our code injection attack inside a Debian environment (Linux Kernel 6.9.6).
Limitations in Rocket Core's performance make it difficult to evaluate our proof-of-concept on a complete runtime environment as the number of processes spawned by a full-blown browser exceed the core's capacity. 
As an alternative analysis, we test our attack using the V8 engine (version 13.0) to process the adversarial JavaScript code.
Figure~\ref{fig:rocket_pipeline} depicts the trojan operation throughout the attack.

\paragraph*{\textbf{Activation Stage}}
The trigger circuit constantly monitors the ALU operands for an IOP activation key.
The JavaScript code accesses selected indices of the injected object, forming this key via the effective address additions generated for each access, as described in Section~\ref{sec:memory_reco}.
Depending on the operand ordering of these additions, only one of the trigger FSMs will reach the final state, recovering the base-pointer of the software payload (operation \circled{1} in Figure~\ref{fig:rocket_pipeline}).
The recovered virtual address and the trigger signal are then forwarded to the payload circuit in the EX stage (operation \circled{2}).

\paragraph*{\textbf{Attack Stage}}
Upon triggering, the payload spoofs a PC-change request in the following clock cycle (operation \circled{3}) through the existing communication bus between the core and the frontend.
Once the next-PC is tainted (operation \circled{4}), the execution flow of the V8 process is diverted and Rocket Core starts fetching instructions from the injected array.
Because V8 heap pages storing JS objects are non-executable, these instruction fetches would normally raise a page fault exception.
Instead, the payload multiplexer forces instruction-related exceptions to zero, suppressing faults detected by the translation lookaside buffer (operation \circled{5}), and allowing injected instructions to proceed to decoding. 

\par We successfully activate the \TROJAN~trojan using both the Ignition interpreter and the TurboFan compiler, bypassing the ASLR mechanism that randomizes V8's address space, and conclude a code injection attack before restoring the V8 execution to its normal control flow.

\section{\TROJAN-Trigger Robustness Analysis}
\label{sec:surf_evaluation}
    \par Native hardware support for integer operations across platforms ensures that their high-level counterparts are consistently lowered to the same hardware primitives.
Hence, we focus our \TROJAN-trigger robustness analysis around effective address calculations, being potentially more prone to variation across code samples and runtime versions. 

\par To evaluate our trigger prototype, we combine our FPGA setup with QEMU-based emulation of V8 engine's operation.
Specifically, we use a RISC-V user-mode QEMU instance, extended with a custom plugin that tracks every machine instruction executed by V8.
The resulting instruction traces provide fine-grained visibility into the engine's software evolution, while the FPGA platform evaluates \TROJAN-trigger effectiveness under realistic operating system conditions.
Both experimental environments use identical V8 binaries and JS workloads. 



\subsection{Detection Resilience}
\label{sec:surf_detection_resilience}
    \par For an attacker, detection resilience determines whether a hardware trojan remains a durable foothold or is rapidly neutralized.
If trigger generation maps to a fixed code signature, defenders can detect and block it at scale.
Conversely, triggers generated using diverse code sequences weaken signature-based defenses by eliminating a stable detection pattern.
Likewise, if defenders can trivially discover the trigger condition, they can deliberately activate and analyze the trojan to accelerate reverse engineering and fleet-wide remediation.
Resilience against both signature-based detection and exhaustive trigger discovery preserves trojan stealthiness, prolongs persistence, and prevents rapid containment.

\paragraph*{\textbf{Evading Signature-based Detection}}
Code polymorphism is a well-known technique for evading signature-based detection~\cite{PSPF2001, ZHZQXZMTZ2005, SYLMESAKADSS2007}, whereby malicious programs maintain semantic equivalence while varying their code structure.
The trigger functions in Figure~\ref{fig:polymorphism_listings} exemplify this property by performing identical memory indexing operations through distinct control-flow constructs.
To quantify their structural similarity at the machine-code level, we execute each variant in V8 under QEMU to collect instruction traces, retaining only the trigger function instructions generated by the Ignition interpreter.

\par We measure similarity using the normalized Levenshtein distance of the instruction mnemonics.
The Levenshtein distance~\cite{YLBL2007} measures the minimum number of single entity edits required to transform one sequence into another.
We compare mnemonics rather than raw instruction bytes to capture structural differences in the V8 source code while disregarding register allocation.
Table~\ref{table:trigger_similarity} shows that the instruction sequences differ substantially despite producing identical effective address calculations. 
Consequently, \textit{IOP generation remains consistent across structurally diverse code samples}, enabling \TROJAN-trigger activation while complicating signature-based detection, analogous to polymorphic malware~\cite{PMAKGMEP2009}.

\lstset{
    basicstyle=\ttfamily\footnotesize,
    backgroundcolor=\color{white},
    numbers=left,
    numberstyle=\tiny\color{gray},
    keywordstyle=\color{blue},
    commentstyle=\color{green!50!black},
    stringstyle=\color{red},
    breaklines=true,
    columns=fullflexible,
}

\begin{figure}[t]
    \centering

    \begin{minipage}{0.9\columnwidth}
    \begin{lstlisting}[caption=Main, language=Java, frame=none]
let fsm_states = 7, stride = 17, i = 0;
let temp = new Array(1);
const object = new Array();
function main(){ trigger(); }
    \end{lstlisting}
    \end{minipage}


    \begin{minipage}{0.9\columnwidth}
    \begin{lstlisting}[caption=For -- Trigger, language=Java, frame=none]
function trigger(){
  for(i = 0; i < fsm_states; i++){
    temp = object[offset_start + i * stride];}}
    \end{lstlisting}
    \end{minipage}


    \begin{minipage}{0.9\columnwidth}
    \begin{lstlisting}[caption=Recursive -- Trigger, language=Java, frame=none]
function recursive(i, max, offset, stride, x, y){
  if (i >= max) return;
  y = x[offset + i * stride];
  recursive(i + 1, max, offset, stride, x, y);}
function trigger(){
  recursive(0, fsm_states, offset_start, stride, object, temp);}
    \end{lstlisting}
    \end{minipage}


    \begin{minipage}{0.9\columnwidth}
    \begin{lstlisting}[caption=While -- Trigger, language=Java, frame=none]
function trigger(){
  offset = offset_start;
  while(i < fsm_states){
    temp = object[offset]; offset += stride; i++;}}
    \end{lstlisting}
    \end{minipage}

    \caption{Semantically equivalent but structurally different
    JavaScript trigger functions for IOP generation.}
    \label{fig:polymorphism_listings}
\end{figure}

\begin{table}[t]
\caption{Structural comparison of different trigger function implementations, using the normalized Levenshtein distance on mnemonics of instruction traces captured in QEMU.}
\footnotesize
\centering
\scalebox{1}{
\begin{tabular}{*{4}{c}}
    \toprule
    \textbf{\shortstack{Language\\Processor}}&
    \textbf{\shortstack{V8\\Version}}&
    \textbf{\shortstack{Structures\\Compared}}& 
    \textbf{\shortstack{Mnemonics\\Similarity (\%)}}\\
    \toprule
    \multirow{3}{*}{Ignition}& \multirow{3}{*}{13.0}& \texttt{For} vs. \texttt{While}& 60\\
    & & \texttt{For} vs. \texttt{Rec}& 37\\
    & & \texttt{While} vs. \texttt{Rec}& 31\\
    \bottomrule
\end{tabular}
}
\label{table:trigger_similarity}
\end{table}

\paragraph*{\textbf{Resisting IOP Exhaustive Search}}
The trigger design in Section~\ref{sec:surf_trigger_design} supports arbitrarily long sequences of integer-operation stimuli, making accidental formation of an IOP key during benign software execution highly improbable.
Targeted activation attempts, however, pose a stronger threat.
We therefore analyze a defender who possesses partial knowledge of the trojan and attempts to expose it.
Following the attack scenario of Section~\ref{sec:end_to_end_attack}, we assume that the defender gained insight to the object size required to generate the relevant memory indexing operations for the IOP.
However, the defender has no knowledge of the index values forming the IOP key.
The defender can thus use this information to exhaustively exercise candidate indices and force the trojan from its dormant state.

\par The time to fully activate the trigger is termed as \textit{trigger time}~\cite{DBLP:YHDAS16}.
A reasonable search strategy is to sequentially access every object element, to ensure all indexes in the predefined set are exercised.
To guarantee activation for any key, the defender must traverse the entire object once for each state of the trigger FSM.
Approximating each effective address calculation as an addition instruction, the worst-case search requires
\begin{equation}
    \text{BF}_{\text{instructions}} = \text{ST} \times \text{ OS }
    \label{equ:bf_adds}    
\end{equation}
where \textit{ST} represents the state transitions and \textit{OS} represents the object size in elements.
This yields a maximum trigger time
\begin{equation}
    \text{BF}_{\text{trigger-time}} = \frac{\text{BF}_{\text{instructions}}}{\text{ IPC }} \times \text{ CT }
    \label{equ:bf_time}
\end{equation}
where \textit{IPC} is the number of instructions retired per cycle and \textit{CT} is the clock period in \text{ns} per cycle.
Under our 3PIP threat model, the attacker aims to maintain processor performance during the trojan integration.
Realistically, to harden the design against IOP brute-force attacks, the adversary can tune both the number of FSM states and the largest predefined index, therefore affecting the size of the object in our attack scenario.

\par We consider a seven-state \TROJAN~FSM implemented in a 4 GHz processor with an IPC of 1.
With an object size of 16 K elements, Equation~\ref{equ:bf_time} yields a maximum trigger time of approximately 24 $\mu s$, providing insufficient resistance to exhaustive search.
To address this, we introduce a new tunable parameter, \textit{Trials}, that regulates the number of integer operations sharing a constant free operand that may occur in selected FSM states.
Exceeding this threshold resets the FSM to its initial state.
The trigger logic can consequently contain both \textit{regular} and \textit{limited} states, which permit at most \text{Trials} operations.
The resulting worst-case instruction count is:
\begin{equation}
    \text{BF}_{\text{instructions}} = (\text{ST}_\text{R} \times \text{ OS } + \text{Trials}) \times (\frac{\text{OS}}{\text{Trials}})^{\text{ST}_\text{L}}
    \label{equ:bf_time_exp}
\end{equation}
where $\text{ST}_\text{R}$ is the number of regular states and $\text{ST}_\text{L}$ the number of limited states, respectively.
This modified reset condition in the number of limited states, makes exhaustive IOP-key search exponentially long.
For reference, limiting the final three states of the seven-state FSM to two integer operations each increases the maximum trigger time to approximately 71 days.
These results show that \TROJAN~trojans can be hardened against exhaustive key-recovery attempts with modest logic changes in the trigger FSM.
To ensure clarity, the FSM implementation in Section~\ref{sec:surf_implementation} incorporates limited states.

\subsection{Operational Resilience}
\label{sec:surf_operational_resilience}
    \begin{table}[t]
\caption{Trigger function's mnemonics similarity using the normalized Levenshtein distance. Mnemonics generated by Ignition across different V8 versions.}
\scriptsize
\centering
\scalebox{1}{
\begin{tabular}{*{8}{c}}
    \toprule 
    \multicolumn{5}{c}{\textbf{Similarity Against V8 10.0 (\%)}}\\[-1em]
    \textbf{10.0}& \textbf{10.7}& \textbf{11.0}& \textbf{12.0}& \textbf{13.0}&
    \textbf{\shortstack{JS\\Code}}&
    \textbf{\shortstack{\TROJAN~\\Activation}}&
    \textbf{\shortstack{Code \\ Injection}}\\
    \toprule
    100 & 91 & 90 & 66 & 50 & \texttt{For} & \checkmark
    & \checkmark\\
    100 & 91 & 90 & 74 & 65& \texttt{Recursive} & \checkmark
    & \checkmark\\
    100 & 92 & 92 & 62 & 48& \texttt{While} & \checkmark
    & \checkmark\\
    \bottomrule
\end{tabular}
}
\label{table:v8_similarity}
\end{table}



\par Operational resilience determines whether an HT can maintain a persistent and controllable foothold in an infected system under realistic operating conditions as the surrounding software stack evolves.
A trigger mechanism tightly coupled to fragile software artifacts may fail under routine updates, recompilation, or structural changes, compromising long-term persistence~\cite{KDJC23}.
Similarly, unrealistic execution assumptions about the trigger software (\textit{i.e.,} uninterrupted control flow~\cite{MAMFKAD24}) can undermine reliable and repeatable activation in the presence of normal operating system behavior.
Designing trigger circuits that tolerate software evolution and OS-level dynamics is therefore essential for maintaining robustness over time.

\paragraph*{\textbf{Software Evolution Resilience}} 
Runtime engines continuously evolve to support the performance and security demands of their host applications.
Beyond updates of individual functions, these systems may also undergo drastic architectural changes that can impact the trojan's activation mechanism.

To evaluate the resiliency of our attack to V8 software evolution, we use QEMU to examine the prevalence of the base-plus-offset addressing strategy relied upon in Section~\ref{sec:surf_detection_resilience} across five distinct V8 versions~\cite{V8releases}.
The versions span a three-year development period, from January 2022 to January 2025.
For each version, we collect the machine instruction mnemonics generated by the V8 engine's language processors during the trigger function execution.
We then use the normalized Levenshtein distance to quantify the mnemonic-sequence similarity across V8 versions.
Table~\ref{table:v8_similarity} presents our findings for Ignition.
Results for Turbofan are nearly identical.

\begin{figure}[t]
    \centering
    \includegraphics[width=0.9\columnwidth]{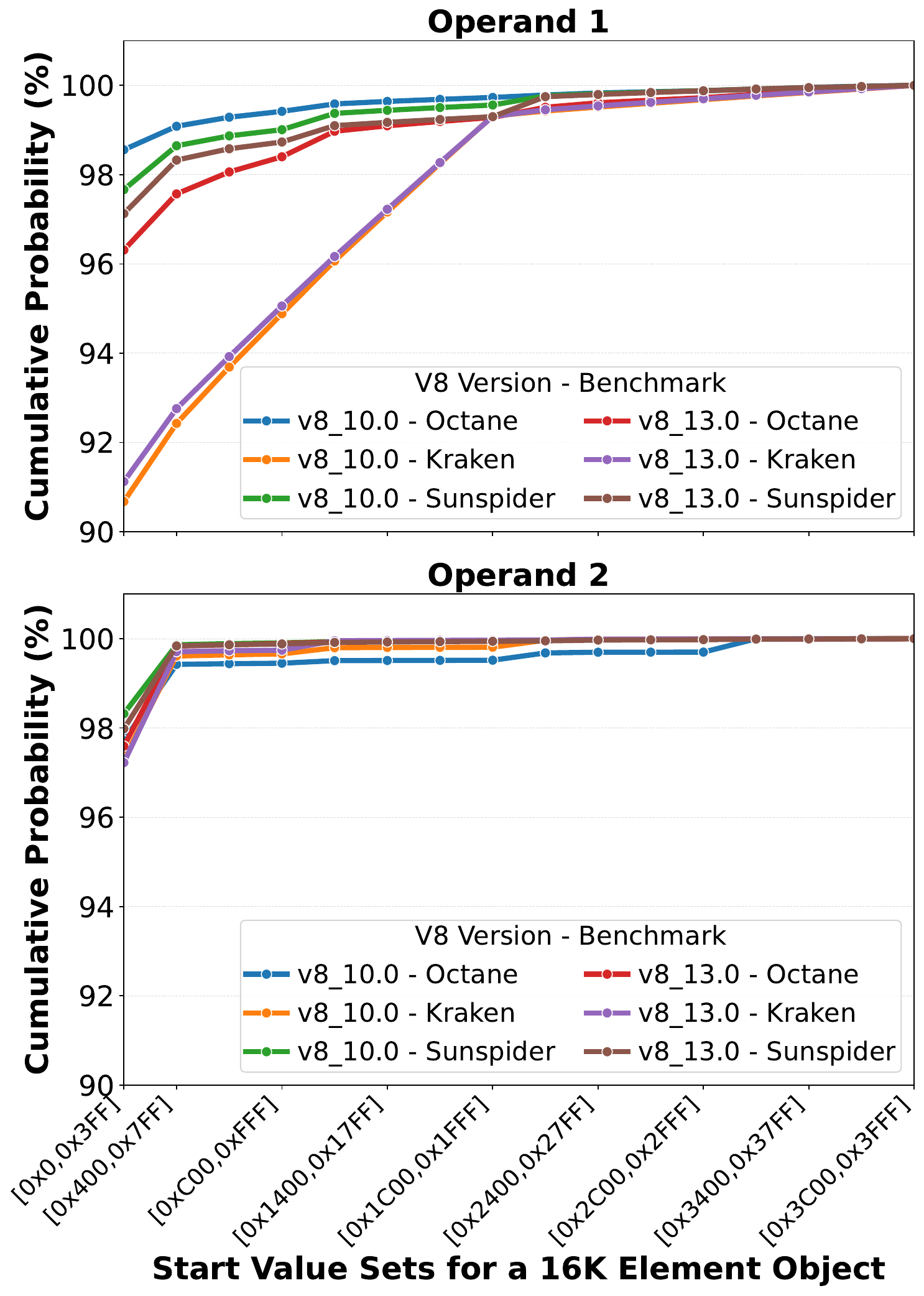}
    \caption{Cumulative probability distribution of IOP start values in Octane, Kraken and Sunspider benchmark suites.}
    \label{fig:operands_cdf}
\end{figure}

\par Measurement-wise, similarity drops sharply after version 11.0, suggesting substantial source code variance.
To identify the origin of this drop amid the thousands of files modified in each release, we collect the symbols associated with each logged instruction in the traces.
Using the instruction population per symbol as our metric, we compare versions 10.0, 12.0 and 13.0, focusing on differences exceeding $30\%$.
The primary contributors are \emph{(i)} additional runtime checks, \emph{(ii)} increased loop-dispatching handling, \emph{(iii)} restructured inline caching and property-lookup for JS object access, and \emph{(iv)} changes to load/store handlers.
Despite these changes, we validate on Rocket Core that \TROJAN~trojans successfully trigger and perform code injection across all tests in Table~\ref{table:v8_similarity}.

\par These findings show that the \TROJAN-enabled code injection attack of Section~\ref{sec:end_to_end_attack} can persist over time, as \textit{IOP generation remains consistent over years of V8 engine evolution}.

\paragraph*{\textbf{Context Switching Resilience}}
Over a CPU's lifetime, benign workloads may inadvertently advance the trigger FSM.
The reset mechanism of Section~\ref{sec:surf_trigger_design} protects \TROJAN-enabled offensives from such transient progressions, but introduces a caveat during the activation stage.
As execution of the HT control software is time-sliced by the OS-enforced context switching, a benign integer operation satisfying the reset condition may abort an offensive midway its activation stage.
This is consistent with prior work highlighting context switching as a challenge for HT offensives~\cite{DBLP:TM14, DBLP:DeKNG20, MAMFKAD24}.
Nevertheless, the activation window is short---determined by the FSM depth, the trigger code, and its runtime realization---relative to the processor uptime, making this risk tolerable.
Importantly, attackers can further minimize the risk by carefully selecting a \say{rare} enough start value for the \texttt{IO:RESET} pair of Figure~\ref{fig:IOP_Trigger_Design}.

\par We note here that rareness in general-purpose computing is inherently workload-dependent.
Accordingly, prior work approximates it empirically through representative benchmarks~\cite{DBLP:YHDAS16,DBLP:KAFP19,ZJZYLHJJ20,ACM:VGHGSPJR22,KDJC23} or heuristics~\cite{KMHHCMLKJ19,DBLP:DeKNG20,MAVKKAD2022,KDJC23,MAMFKAD24}.

\par We evaluate the resilience of our trigger circuit---and consequently of the code injection attack---to context switching interference.
Our methodology consists of two stages.
First, we explore from the \TROJAN-trigger's point of view, the existence of rare start values in runtime-related traffic.
Second, we select several of these rare start values to test the attack's robustness under traffic interference on the ALU inputs from multiple endpoint workloads.
We perform the relevant resiliency experiments on the Rocket Core FPGA implementation.

\par To identify rare start value candidates for the \texttt{IO:RESET} pair used in our attack, we generate background traffic across the three main operational axes of a JavaScript runtime: just-in-time compilation, garbage collection, and runtime execution.
For this purpose, we use the widely adopted Sunspider, Octane, and Kraken JavaScript benchmark suites, which subject V8 to sustained computations beyond typical browser activity.
Our setup counts ALU operand pairs whose free operand exhibits the address-like form described in Section~\ref{sec:surf_implementation}.
We restrict the static operand start value to the range \texttt{0}--\texttt{0x3FFF} (0--16383), for two reasons:
\emph{(i)} address-related integer operations may heavily populate values up to \texttt{0xFFF}, as the RISC-V  addressing modes extensively use the 12-bit immediate field for effective address calculations, and \emph{(ii)} extending the range to \texttt{0x3FFF} provides sufficient room to move beyond this dense region while keeping the object size required by our attack practical and inconspicuous.

\par We execute each suite for its recommended number of iterations on V8 versions 10.0 and 13.0, with individual measurements lasting up to 17 hours.
Start values are grouped in sets of 1024 values, yielding a total of 16 sets.
Because JavaScript does not control which ALU input receives each IOP operand, our setup tracks both operand orderings.
Figure~\ref{fig:operands_cdf} shows the cumulative distribution of the measured start values.

\par The results reveal a strongly skewed distribution for integer operations pairing address-like values with small operands.
More than $90\%$ of these small operands fall within the lowest 10 bits (\textit{i.e.,} at or bellow \texttt{0x3FF}), while $99\%$ lie below the \texttt{0x2000} threshold.
This motivates selecting an infrequent start value just beyond \texttt{0x2000}, where benign occurrences become substantially less likely while the object size remains practical.

\par We empirically validate our findings by measuring the trigger success rate for eight start values above \texttt{0x2000} while concurrently executing varying workloads.
Representative endpoint activity is hard to model because it spans highly heterogeneous resource demands.
We therefore combine the above JS benchmark suites with the Phoronix Test Suite~\cite{Phoronix2026}.
The latter provides standardized Linux benchmarks representative of common endpoint resource usage:
storage and database activity with FIO and SQLite benchmarks; network traffic with iPerf; memory and cache stress workloads with MBW and Multichase; program launching and scheduling with OSBench and Hackbench; and compute-intensive cryptographic, compression, and image decoding workloads via Stress-NG.

\par We use the default configuration of each selected benchmark and randomly select two start values from each of the sets 9, 10, 11 and 13, to experiment on Rocket Core.
Each experiment includes the parallel execution of two processes.
The first sequentially executes the above benchmarks spanning a period of approximately 4 hours.
During this time period, the second process randomly selects one of the trigger functions in Figure~\ref{fig:polymorphism_listings}, starts the V8 engine, and executes the selected code using either Ignition or Turbofan.
Since the \TROJAN~trojan uses the same IOP to alternate between the enable and disable state, we execute the trigger function twice.  
The process then sleeps for a few seconds before repeating the cycle of random selection and execution for a total of 1600 times, yielding 3200 total trigger toggles per start value.
The results across all tested offsets show a trigger success rate that exceeds $99\%$.
We argue that \textit{IOPs remain highly resilient to context switching events}, enabling reliable activation and deactivation of \TROJAN~trojans under realistic operating conditions.

\section{Software-based Defenses}
\label{sec:sw_based_defenses}
    \par Our threat model assumes a processor 3PIP containing a \TROJAN~trojan is integrated into a chip, passes standard security evaluations, and reaches the market as a COTS product.
Trust verification in deployed COTS devices is particularly challenging~\cite{HMCJCPBSHT2022}, as many trojan defenses rely on assumptions unavailable in this setting.
Approaches requiring hardware design modifications (\textit{e.g.,} sensor implants~\cite{CYCCHCS2014}), white-box access to the design IP~\cite{CRSWFPSPCBW2009, LEJYMY2012} or golden reference models (\textit{e.g.,} side-channel signatures~\cite{DBLP:NguyenCPZ19}) are unsuitable on a finished product.
Processing element redundancy~\cite{BMHBNT2012} is likewise impractical for cost-sensitice end-user devices.
In practice, manufacturers often mitigate hardware faults through software, even at the expense of performance~\cite{KPHJFAGDGDHWHMLMMSPTSMYY2019}.
Because \TROJAN~trojans target end-user devices at scale, runtime detection and mitigation is particularly important, as product recalls can be prohibitively expensive~\cite{ErSm11, MiIs11}.
We therefore focus on software-based defenses~\cite{HMCJCPBSHT2022, MASESGJMUIAMSMFMPPDSASM2018} that require no hardware modifications.

\par The work of~\citet{HMCJCPBSHT2022} aims at trojan detection via software execution redundancy.
Their approach partitions programs into code blocks, generates structurally different but semantically equivalent variants, and executes them to induce different internal switching activity that may expose HT behavior.
However, their approach is vulnerable to trojans triggered through diverse code variants.
This renders it ineffective against \TROJAN~trojans, which can be activated by different integer instructions and structurally dissimilar code as shown in Tables~\ref{table:trigger_similarity}~and~\ref{table:v8_similarity}.
Moreover, their method targets payloads that induce transient data corruption, and may therefore miss attacks such as ours that restore the original program state.

\par In a similar direction,~\citet{MASESGJMUIAMSMFMPPDSASM2018} prevent trojan activation via code obfuscation.
Their technique applies four functionality-preserving assembly-level mutations to eliminate instruction patterns used as trigger sequences, including those of the A2 trojan~\cite{DBLP:YHDAS16}.
However, the authors note their method is ineffective against trigger circuits that monitor values on data paths, which is the principle our trigger operates on.



\section{\TROJAN-Trigger (Micro)architectural Portability}
\label{sec:surf_portability}
    \begin{table}[t]
\caption{\TROJAN~Stand-Alone Metrics.}
\scriptsize
\centering
\scalebox{1}{
\begin{tabular}{c c c c c c}
    \toprule
    \textbf{Design}&\textbf{\shortstack{\# Comb.\\Cells}}&\textbf{\shortstack{\# Seq.\\Cells}}&\textbf{\shortstack{Frequency\\(GHz)}}&\textbf{\shortstack{Static\\Leakage (mW)}} &\textbf{\shortstack{Dynamic\\Power (mW)}}\\
    \midrule
    \TROJAN& 518& 100& 2.0&  0.35& 4.32\\
    \bottomrule
\end{tabular}
}
\label{table:surf_d_stats}
\end{table}

\par A portable \TROJAN-trigger provides a reusable primitive that can target multiple microarchitectures with limited modification.
Rocket Core features a single ALU, unlike industry-grade RISC-V~\cite{CCXXLCSYGRLDLYHZLJCZLCPYMJYZXYQX2020, WW2025, LC2025} and x86~\cite{WCAMDZEN2, AMDREFMANUAL} superscalar processors.
Superscalar execution, however, does not fundamentally constrain a design stage attacker: the trigger logic can be vectorized to track operand pairs across ALUs, letting a shared FSM to advance on their aggregated result.

\par A key concern is the simultaneous processing of multiple IOP elements in the same cycle, since the FSM advances by only one state per cycle.
To avoid this, IOP elements must reach ALUs in separate cycles.
This separation arises naturally in JavaScript due to the intermediate instructions added by runtime checks.
For example, lowering the \texttt{For} function in Figure~\ref{fig:polymorphism_listings} under V8 version 13.0 places more than 2000 instructions between consecutive IOP-relevant calculations, allowing the FSM to maintain correct pattern recognition.
Likewise, this large spacing mitigates out-of-order effects, allowing IOP instructions to mirror their program order upon execution.

\par Consider also that RISC-V is a load-store architecture, performing address calculations via integer execution units.
In contrast, register-memory architectures like x86~\cite{WCAMDZEN2, RWTSANDYBRIDGE, INTELREFMANUAL, AMDREFMANUAL} delegate this calculation to address generation units or AGUs.
Thus, an adversary will target the AGUs to track the IOP.

\par To verify that V8 preserves the same base-plus-index addressing strategy across architectures, we execute the trigger functions in Figure~\ref{fig:polymorphism_listings} on an x86 build of V8 on QEMU.
Ignition and Turbofan traces show that array accesses use \texttt{movzbl} instructions with indexed addressing.
The effective address is computed through an AGU adding two architectural registers containing the array base pointer and the accessed index.
Thus, the activation principle underlying our strategy in Section~\ref{sec:memory_reco} extends naturally to x86-class processors.

\par Finally, we synthesize the \TROJAN~trojan we used in our code injection attack on a 28nm node process using the Genus tool from Cadence.
The synthesis results in Table~\ref{table:surf_d_stats} indicate that its timing is compatible with clock frequencies reported by commercial x86 processors at the same process node~\cite{WCCFTECHISIAH2}.







\section{Conclusions}
\label{sec:conclusions}
    \par This work introduces \TROJAN, a new class of hardware trojans that can be reliably activated without requiring arbitrary code execution on the victim CPU.
To achieve that, \TROJAN-trigger circuits leverage the predictable hardware manifestation of integer operations that can be expressed in high-level code and reach the victim system through runtime systems prevalent in modern software stacks.
We prototype a \TROJAN~trojan inside a RISC-V processor and demonstrate a code-injection attack that hijacks the control flow of the V8 JavaScript engine in a Linux environment, enabling arbitrary machine code execution.

\par Our analysis shows that \TROJAN-triggers can evade signature-based defenses through diverse activation stimuli, resist exhaustive key search attempts, and remain robust to context switching events and runtime software updates.
Together, these properties allow \TROJAN~trojans to maintain a durable and evasive foothold, enabling prolonged offensive campaigns that target end-user devices. 
To the best of our knowledge, this work is the first to systematically address the practical challenges of activating CPU hardware trojans via runtime engines, demonstrating their threat to end-user devices.


\bibliographystyle{IEEEtranN}
\bibliography{references}

\end{document}